# Describing the heavy-ion above-barrier fusion using the bare potentials resulting from Migdal and M3Y double-folding approaches

I. I. Gontchar[1], M. V. Chushnyakova[2,3]

[1]*Physics and Chemistry Department, Omsk State Transport University, Omsk, Russia*
[2]*Physics Department, Omsk State Technical University, Omsk, Russia*
[3]*Applied Physics Department, Tomsk Polytechnic University, Tomsk, Russia*

E-mail: maria.chushnyakova@gmail.com

**Abstract**
Systematic calculations of the Coulomb barrier parameters for collisions of spherical nuclei are performed within the framework of the double folding approach. The value of the parameter $B_Z = Z_P Z_T/(A_P^{1/3} + A_T^{1/3})$ (which estimates the Coulomb barrier height) varies in these calculations from 10 MeV up to 150 MeV. The nuclear densities came from the Hartree-Fock calculations which reproduce the experimental charge densities with good accuracy. For the nucleon-nucleon effective interaction two analytical approximations known in the literature are used: the M3Y and Migdal forces. The calculations show that Migdal interaction always results in the higher Coulomb barrier. Moreover, as $B_Z$ increases the difference between the M3Y and Migdal barrier heights systematically increases as well. As the result, the above barrier fusion cross sections calculated dynamically with the M3Y forces and surface friction are in agreement with the data. The cross sections calculated with the Migdal forces are always below the experimental data even without accounting for the dissipation.

*Keywords:* Heavy-ion fusion; Double folding potential; Migdal and M3Y *NN* forces

## 1. Introduction

Potential energy of the nuclear part of nucleus-nucleus interaction is the key ingredient for any theoretical description of fusion of two complex nuclei [1]. For this energy Woods-Saxon profile is often used [2-6]:

$$U_n(R) = V_{WS}\left\{1 + \exp\left(\frac{R - r_{WS}(A_P^{1/3} + A_T^{1/3})}{a_{WS}}\right)\right\}^{-1}. \tag{1}$$

Here $R$ stands for the center-to-center distance of the colliding nuclei (only spherical nuclei are considered in the present work), $A_P$ ($A_T$) is the mass number of the projectile (target) nucleus. This profile is characterized by three parameters, namely, the depth $V_{WS}$, the radius $r_{WS}$, and diffuseness $a_{WS}$. Their values are varied more or less arbitrary in order to make the calculated fusion cross sections fitting the experimental ones at the collision energies well above the barrier. We think this profile represents the real nucleus-nucleus interaction only qualitatively.

The proximity potential [7] used e.g. in Refs. [8-10] seems to be much better founded. This potential reads:

$$U_n(R) = 4\pi\gamma b P_{sph}^{-1} \Phi\left(\frac{\xi}{b}\right). \tag{2}$$

Here $\Phi\left(\frac{\xi}{b}\right)$ is an universal dimensionless function, $b$ is of order of 1 fm and represents the thickness of the surface, $\gamma = \gamma_0(1 - 1.7826 I^2)$, $\gamma_0 = 0.95$ MeV fm$^{-2}$ is the surface tension coefficient of the nuclear liquid, $I = (N-Z)/(N+Z)$, $\xi$ is the distance between the surfaces of the interacting nuclei, $P_{sph} = \overline{R}_P^{-1} + \overline{R}_T^{-1}$, $\overline{R}_{P(T)} = R_{P(T)}(1 - b^2 R_{P(T)}^{-2})$. The proximity potential includes parameters $R_P, R_T, b$ which in principle can be varied individually for each reaction.

The next level of generality is represented by the single-folding potential [11, 12]

$$U_n(R) = \frac{1}{2}\left\{\int d\vec{r}_P \rho_{AP}(\vec{r}_P) V_T(\vec{R} - \vec{r}_P) + \int d\vec{r}_T \rho_{AT}(\vec{r}_T) V_P(\vec{R} - \vec{r}_T)\right\}. \tag{3}$$

Here $\rho_{AP}(\vec{r}_P)$ [$\rho_{AT}(\vec{r}_T)$] is the distribution of the nucleon centers of mass (the nucleon density) for the projectile (target) nucleus, $V_T(\vec{R} - \vec{r}_P)$ [$V_P(\vec{R} - \vec{r}_T)$] stands for the energy of the interaction between a nucleon of the projectile (target) nucleus with the whole target (projectile) nucleus. For the nucleon-nucleus potential usually the Woods-Saxon profiles (1) are used with the parameters extracted from the fit of the elastic scattering data. Thus for given reaction one has six individual fit parameters for the nucleus-nucleus potential. For the nuclear densities in Refs. [11, 12] the two-parameter Fermi-profile was used:

$$\rho_A(r) = \rho_0 \left\{ 1 + \exp\left(\frac{r - R_A}{a_A}\right) \right\}^{-1}. \tag{4}$$

The normalization constant $\rho_0$ was found from the nucleon number conservation, whereas the radius and diffuseness parameters, $R_A$ and $a_A$, might be extracted from the electron scattering experimental data [13]. However one should keep in mind that the electron scattering is only sensitive to the Coulomb interaction. Therefore in such experiments the charge density distribution $\rho_q$ is measured rather than the nucleon density $\rho_A$. No procedure exists presently for finding $\rho_A$ when $\rho_q$ is known. Thus in the single-folding approach there are some problems and no ways are seen for solving them.

The next step towards a more realistic description of the nucleus-nucleus potential is to average the effective nucleon-nucleon forces accounting for the density distribution in both interacting nuclei. Two such double-folding (DF) potentials are known in the literature: the one using M3Y nucleon-nucleon forces [14, 15] and another with the Migdal forces [16]. The aim of the present work is to compare systematically the heavy-ion Coulomb barriers obtained within the framework of the DF approach with these two different *NN*-interactions and to see what are the resulting fusion cross sections. It should be noted that this approach operates within the frozen density approximation which is supposed [17] to produce the first-order term of the real part of the microscopic optical potential.

Presently more sophisticated time-dependent Hartree-Fock (TDHF) calculations [18 - 23] of the heavy-ion collisions are available. This approach is extremely computer time demanding and provides only deterministic trajectory without fluctuations. However it is very useful for establishing the borders of applicability of simpler models like the ones used in the present work.

Within the TDHF approach the self-consistent evolution of the nucleon density profile and nucleus-nucleus interaction is obtained. In these works it was shown that the nucleon transfers [19] as well as the dynamical reagents deformation and neck formation [20] might be of importance. Although the probability of the nucleon transfer is known to increase with the collision energy, its effect on fusion is less important above the barrier [19]. In our calculations [24] for the reactions $^{16}$O+$^{92}$Zr, $^{16}$O+$^{144}$Sm, $^{16}$O+$^{208}$Pb capture was decided when the density in the overlap region was less than 25% of its central value (see, e.g. Figure 9 of Ref. [25]). Thus, for the reactions considered in the present work, the neck certainly appears and the reagents deform but after the capture is decided and therefore beyond the framework of our model.

## 2. The double-folding models

### 2.1. M3Y double-folding-potential

The DF potential with M3Y forces was first systematically used in [17] in order to calculate the real part of the microscopic optical potential for describing elastic and inelastic scattering of alpha-particles and heavy ions. Later this potential was used in many works [26-30] for the same goal but with the modified M3Y *NN*-forces. Comparatively recently this potential started to be applied for the fusion problem [24, 25, 31-34]. In Refs. [24, 25, 32] the M3Y DF potential was discussed in relation with the so-called "problem of large diffuseness" [3] which can be formulated as follows.

Systematic analysis of the experimental capture excitation functions in Ref. [3] demonstrated that the diffuseness of the potential ranging between 0.75 and 1.5 fm were needed to reproduce the above barrier parts of those functions. This was much larger than the value of 0.65 fm, which was required by the elastic scattering data. Note that the analysis of the above barrier fusion cross sections in [3] was performed within the framework of the single barrier penetration model. It was pointed out in Ref. [3] that these abnormally large diffusenesses might be an artifact masking some dynamical effects. Following this idea, we analyzed in Ref. [24] the above barrier experimental capture excitation functions using a dynamical model accounting for dissipation of the collision energy and fluctuations of the collective momentum. In this model, the DF potential with M3Y forces was the crucial ingredient. Results obtained in [24] demonstrated the problem of the apparently large diffuseness of the potential was indeed an artifact related to the dynamics of the process.

Detailed description of the M3Y DF nucleus-nucleus potential can be found in many works (see, e.g., [26, 35, 36]), therefore we present here only basic formulas. According to the general rules of quantum mechanics the potential consists of the direct $U_{nD}$ and exchange $U_{nE}$ parts:

$$U_{n\,\text{M3Y}}(R, E_P) = U_D(R, E_P) + U_E(R, E_P). \tag{5}$$

The direct part reads:

$$U_D(R, E_P) = g(E_P) \int d\vec{r}_P \int d\vec{r}_T \rho_{AP}(\vec{r}_P) F_\nu(\rho_{FA}) v_D(s) \rho_{AT}(\vec{r}_T). \tag{6}$$

Here $\vec{s} = \vec{R} + \vec{r}_T - \vec{r}_P$ corresponds to the distance between two points in the interacting nuclei, $v_D$ is the direct part of the effective NN-forces, the multiplier $g(E_P)$ in our case is very close to unity. The nucleon density $\rho_A(\vec{r})$ is the sum of the proton $\rho_p(\vec{r})$ and neutron $\rho_n(\vec{r})$ densities.

The function $F_v(\rho_{FA})$ is responsible for the density dependence of the NN-forces; it is taken from [35]:

$$F_v(\rho_{FA}) = C_v \{1 + \alpha_v \exp(-\beta_v \rho_{FA}) - \gamma_v \rho_{FA}\}. \tag{7}$$

The density at the middle-point between the centers of two nuclei is used for the argument of this function:

$$\rho_{FA} = \rho_{AP}(\vec{r}_P + \vec{s}/2) + \rho_{AT}(\vec{r}_T - \vec{s}/2). \tag{8}$$

The exchange part looks more complicated:

$$U_E(R, E_P) = g(E_P) \int d\vec{r}_P \int d\vec{r}_T \rho_{AP}(\vec{r}_P; \vec{r}_P + \vec{s}) \times \\ \times F_v(\rho_{FA}) v_E(s) \rho_{AT}(\vec{r}_T; \vec{r}_T - \vec{s}) \exp(i\vec{k}_{rel} \vec{s} m_n / m_R). \tag{9}$$

The expansion method of Refs. [37, 38] is used in order to find the non-diagonal component of the density matrix, $\rho_{AP}(\vec{r}_P; \vec{r}_P + \vec{s})$ and $\rho_{AT}(\vec{r}_T; \vec{r}_T + \vec{s})$:

$$\rho_A(\vec{r}; \vec{r} + \vec{s}) \approx \rho_A(\vec{r} + \vec{s}/2) \hat{j}_1(k_F(\vec{r} + \vec{s}/2) \cdot \vec{s}), \tag{10}$$

$$\hat{j}_1(x) = 3[\sin(x) - x\cos(x)]/x^3. \tag{11}$$

The effective Fermi-momentum $k_F$ is calculated within the framework of the extended Thomas-Fermi approach [38].

The direct $v_D$ and exchange $v_E$ parts of NN-interaction consist of the Yukawa-type terms:

$$v_D(s) = \sum_{i=1}^{3} G_{Di}[\exp(-s/r_{vi})]/(s/r_{vi}), \tag{12}$$

$$v_E(s) = \sum_{i=1}^{3} G_{Ei}[\exp(-s/r_{vi})]/(s/r_{vi}). \tag{13}$$

We use here the values of $r_{vi}$, $G_{Di}$, and $G_{Ei}$ corresponding to [15] (the so-called Paris-forces).

2.2. Migdal double-folding potential

This nucleus-nucleus potential reads

$$U_{nMIG}(R) = \int d\vec{r}_P \int d\vec{r}_T [\rho_{Pn}(\vec{r}_P) v_{nn}(s) \rho_{Tn}(\vec{r}_T) + \rho_{Pp}(\vec{r}_P) v_{pp}(s) \rho_{Tp}(\vec{r}_T) + \\ + \rho_{Pn}(\vec{r}_P) v_{np}(s) \rho_{Tp}(\vec{r}_T) + \rho_{Pp}(\vec{r}_P) v_{pn}(s) \rho_{Tn}(\vec{r}_T)] \tag{14}$$

(see, e.g., [1, 39, 40] and references therein). The components of nucleon-nucleon interaction can be written as follows

$$\begin{aligned} v_{nn}(\vec{r}_P, \vec{r}_T) &= \left[a + 2(g-a)\frac{\rho_{Pn}(\vec{r}_P) + \rho_{Tn}(\vec{r}_T)}{\rho_{Pn}(0) + \rho_{Tn}(0)}\right] \delta(\vec{s}) \\ v_{pp}(\vec{r}_P, \vec{r}_T) &= \left[a + 2(g-a)\frac{\rho_{Pp}(\vec{r}_P) + \rho_{Tp}(\vec{r}_T)}{\rho_{Pp}(0) + \rho_{Tp}(0)}\right] \delta(\vec{s}) \\ v_{np}(\vec{r}_P, \vec{r}_T) &= \left[\varphi + 2(\gamma-\varphi)\frac{\rho_{Pn}(\vec{r}_P) + \rho_{Tp}(\vec{r}_T)}{\rho_{Pn}(0) + \rho_{Tp}(0)}\right] \delta(\vec{s}) \\ v_{pn}(\vec{r}_P, \vec{r}_T) &= \left[\varphi + 2(\gamma-\varphi)\frac{\rho_{Pp}(\vec{r}_P) + \rho_{Tn}(\vec{r}_T)}{\rho_{Pp}(0) + \rho_{Tn}(0)}\right] \delta(\vec{s}) \end{aligned} \tag{15}$$

Here $a = C(f_{ex} + f'_{ex})$, $g = C(f_{in} + f'_{in})$, $\varphi = C(f_{ex} - f'_{ex})$, $\gamma = C(f_{in} - f'_{in})$. The values of the constants are: $C = 300$ MeV fm$^3$, $f_{in} = 0.09$, $f'_{in} = 0.42$, $f_{ex} = -2.59$, $f'_{ex} = 0.54$. This potential is defined by the amplitude of the interaction of nucleons (i) from the peripheral parts of the density distributions, $f_{ex} \pm f'_{ex}$, (i.e. at $\rho_P(\vec{r}_P) + \rho_T(\vec{r}_T) \ll \rho_P(0) + \rho_T(0)$); (ii) from the peripheral part and from the inner part of the density distributions $f_{in} \pm f'_{in}$ (i.e. at $\rho_P(\vec{r}_P) \ll \rho_P(0)$, $\rho_T(\vec{r}_T) \lesssim \rho_T(0)$ and $\rho_P(\vec{r}_P) \lesssim \rho_P(0)$, $\rho_T(\vec{r}_T) \ll \rho_T(0)$); and (iii) from the inner parts of the density distributions $2(f_{in} \pm f'_{in}) - (f_{ex} \pm f'_{ex})$ (i.e. at $\rho_P(\vec{r}_P) \lesssim \rho_P(0)$, $\rho_T(\vec{r}_T) \lesssim \rho_T(0)$).

One can ask how the exchange term is treated in Eq. (14). Neither in the book of Migdal [16], nor in the papers using the DF Migdal potential [1, 39, 40] there is no explicit separation of the effective *NN* forces (Eq. (15)) into direct and exchange parts. According to [16], both direct and exchange parts are mostly accounted for in the empirical constants $f, f'$.

*2.3. Nuclear densities*

In order to find the M3Y and Migdal DF potentials we still have to define the nuclear densities. We use the results obtained in [41] within the framework of the Hartree-Fock approach with the SKX-Skyrme forces accounting for the tensor part. These calculations, which are described in details in Refs. [42, 43], result in the charge density distributions which are in good agreement with the experimental data. This is seen in table 1 where the relative deviation of the calculated rms charge radius from the experimental one, $\zeta_q = 1 - R_{qth}/R_{qexp}$, is presented. One can see that mostly this deviation is larger than the experimental error which is also shown in table 1. However $\zeta_q$ is still rather small: only in two cases it is larger than 0.5%. We did not manage to find in the literature a similar comparison with better agreement. These calculated rms charge radii with SKX Skyrme forces are significantly closer to the data, especially for the light nuclei, than the ones obtained earlier in [44] with the SKP Skyrme forces.

**Table 1.** Deviation of the calculated rms charge radius $R_{qth}$ from the experimental one $R_{qexp}$ [45] as well as experimental error.

| Nucleus | $^{12}$C | $^{16}$O | $^{28}$Si | $^{32}$S | $^{36}$S | $^{92}$Zr | $^{144}$Sm | $^{204}$Pb | $^{208}$Pb |
|---|---|---|---|---|---|---|---|---|---|
| $\zeta_q$, % | -1.99 | -1.64 | -0.29 | 0.13 | -0.06 | 0.48 | 0.06 | 0.15 | 0.19 |
| $\Delta R_{qexp}/R_{qexp}$, % | 0.09 | 0.20 | 0.08 | 0.06 | 0.06 | 0.02 | 0.13 | 0.01 | 0.02 |

In figure 1 the calculated $\rho_{ch\,th}$ and experimental $\rho_{ch\,exp}$ charge distributions are compared by means of the fractional difference

$$\delta = 2(\rho_{ch\,th} - \rho_{ch\,exp})/(\rho_{ch\,th} + \rho_{ch\,exp}). \tag{16}$$

The experimental distribution is obtained using the Fourier-Bessel series [13]

$$\rho_{ch\,exp}(r) = \sum_n a_{FBn} j_0(\pi n r / R_{FB}) \tag{17}$$

for $r < R_{FB}$ and $\rho_{ch\,exp}(r) = 0$ otherwise. The values of coefficients $a_{FBn}$ and cutoff parameters $R_{FB}$ are taken from [13]. Note, that the experimental errors for these coefficients and parameters are not available in [13]. One sees in figure 1 that the calculated charge distributions are in good agreement with the data, too. Thus we hope that the proton and neutron density distributions obtained in the same calculations also represent the real situation.

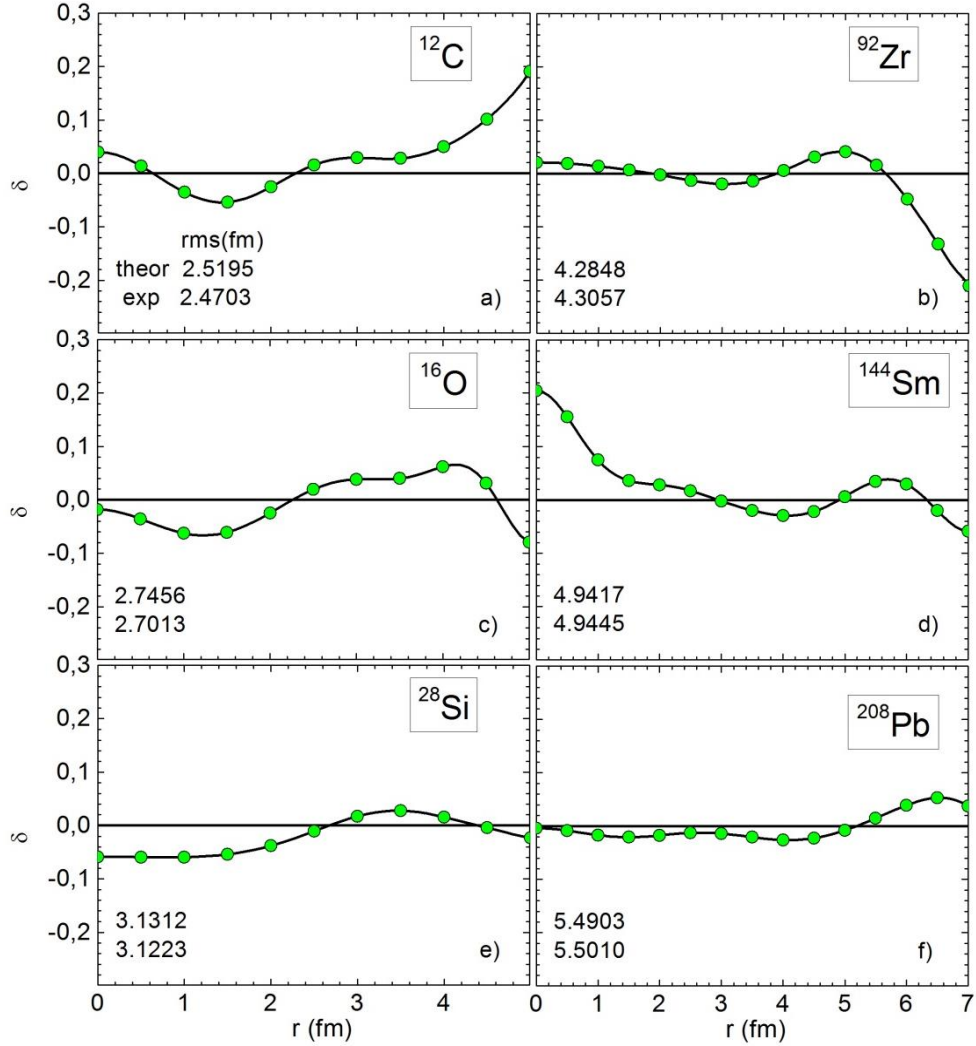

**Figure 1.** (Color online) Fractional difference $\delta$ defined by Eq. (16) for the charge densities versus the distance from the center of the nucleus for some nuclei involved in the present work. The experimental data are taken from [13] (see text for details).

## 3. Comparing the barrier parameters

There are different options for the density dependent M3Y-forces in the literature. Therefore we first show in figure 2 the heights of the Coulomb barriers obtained with different versions of the density dependence for $^{28}$Si+$^{208}$Pb and $^{12}$C+$^{32}$S reactions. The correspondence between the values of coefficients in the density dependence of the M3Y-forces [35] (Eq. (7)) and notations used in figure 2 is presented in table 2. To get rid of the trivial dependence of the charge numbers in figure 2, we divide the barrier height calculated for the zero angular momentum $U_{B0\,\text{M3Y}}$ by the parameter

$$B_Z = Z_P Z_T / \left( A_P^{1/3} + A_T^{1/3} \right) \tag{18}$$

which can be considered as the rough estimate of the Coulomb barrier height. Behavior of the $U_{B0\,\text{M3Y}}$ observed in figure 2 is typical for all the reactions; same behavior was demonstrated earlier in [32] but with the different type of the nucleon density distribution. Note, that the $k_{DD} = -1$ version results in the highest barrier whereas in the case of $k_{DD} = 1 \div 2$ the barriers are the lowest and hardly distinguishable. Thus we choose the zero-range and $k_{DD} = 1$ versions of the M3Y DF potential to be compared with the Migdal DF potential hereafter.

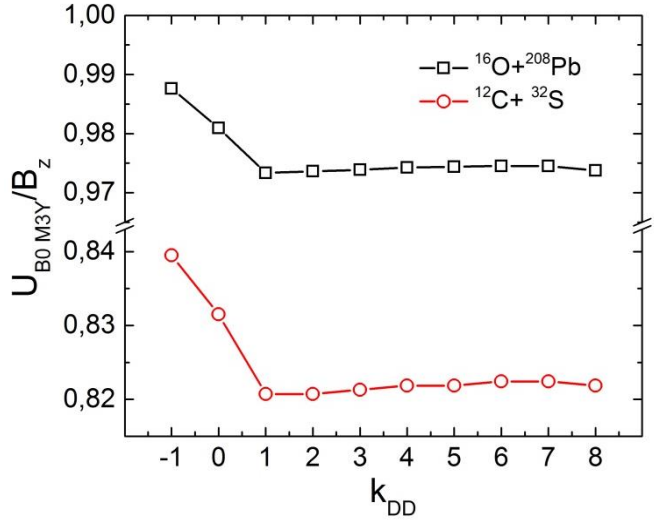

**Figure 2.** (Color online) Variation of the M3Y DF fusion barrier height depending on the version of density dependence (function $F_v(\rho_{FA})$ in Eq. (7)) for two reactions.

**Table 2.** Coefficients of the density dependence of the M3Y-forces [35] (Eq. (7)) and notations used in figure 2. $k_{DD} = -1$ corresponds to the zero-range exchange forces where the density dependence is not applied.

| $k_{DD}$ | $C_v$ | $\alpha_v$ | $\beta_v$ (fm$^3$) | $\gamma_v$ (fm$^3$) |
|---|---|---|---|---|
| -1 (zero range) | 1 | 0 | 0 | 0 |
| 0 (finite range) | 1 | 0 | 0 | 0 |
| 1 | 0.2963 | 3.7231 | 3.7384 | 0 |
| 2 | 0.3429 | 3.0232 | 3.5512 | 0.5 |
| 3 | 0.3346 | 3.0357 | 3.0685 | 1.0 |
| 4 | 0.2985 | 3.4528 | 2.6388 | 1.5 |
| 5 | 0.3052 | 3.2998 | 2.3180 | 2.0 |
| 6 | 0.2728 | 3.7367 | 1.8294 | 3.0 |
| 7 | 0.2658 | 3.8033 | 1.4099 | 4.0 |
| 8 | 1.2521 | 0 | 0 | 1.7452 |

The center of mass distance dependence of these three nucleus-nucleus potentials are presented in figure 3 for the reaction $^{16}$O+$^{208}$Pb. The distinct features of the potentials are that the Migdal DF potential possesses a pocket whereas the M3Y DF potentials do not. Sometimes (see, e.g. [30, 46]) the absence of a pocket in the M3Y DF potentials is considered as their shortcoming. We think that any nucleus-nucleus potential calculated with frozen densities (diabatic potential) is applicable only for rather large center of mass distance $R > R_P + R_T$ (this is explicitly stated in [12]). Therefore it seems irrelevant whether a potential possesses a pocket or not. For all the reactions considered in the present work the condition $R_{B0} > R_P + R_T$ is fulfilled (see table 3 below). Here $R_{B0}$ is the barrier radius at zero angular momentum.

One sees in figure 3 that the barrier of the Migdal DF potential is higher than the highest of the M3Y DF barriers. This feature turns out to be common for all the reactions we consider as seen in figure 4 and in table 3. In figure 4 the fractional differences of the barrier height

$$\xi_U = \left(U_{B0\ \text{MIG}} - U_{B0\ \text{M3Y}}\right)/U_{B0\ \text{MIG}}, \quad (19)$$

radius

$$\xi_R = \left(R_{B0\ \text{MIG}} - R_{B0\ \text{M3Y}}\right)/R_{B0\ \text{MIG}}, \quad (20)$$

and barrier stiffness (frequency)

$$\xi_\Omega = \left(\Omega_{B0\ \text{MIG}} - \Omega_{B0\ \text{M3Y}}\right)/\Omega_{B0\ \text{MIG}} \quad (21)$$

are presented as functions of the parameter $B_z$. For the M3Y DF potential two versions corresponding to the zero-range exchange forces (subscript 'z') and with the density dependent ($k_{DD} = 1$) exchange forces (subscript '1') are used.

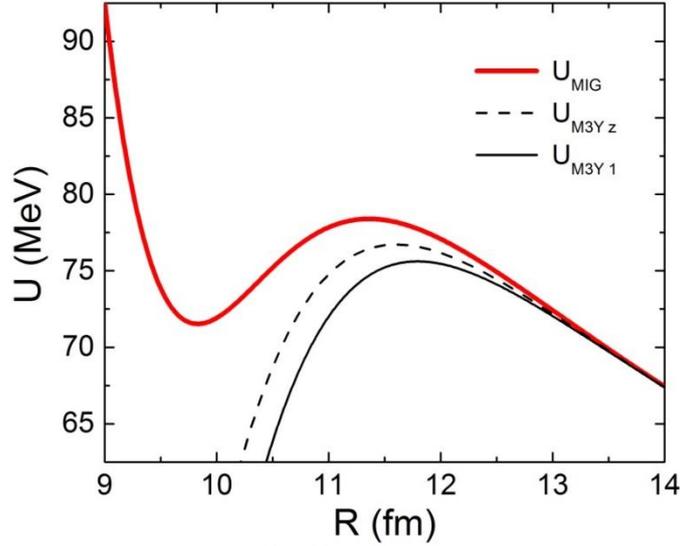

**Figure 3.** (Color online) The nucleus-nucleus potentials for $^{16}$O+$^{208}$Pb reaction versus center of mass distance: Migdal DF ($U_{MIG}$), M3Y-zero-range DF ($U_{M3Yz}$), M3Y density dependent DF ($k_{DD} = 1, U_{M3Y1}$).

The Migdal DF barriers are always higher by several percent (panel a) whereas the barrier radii are smaller. Such type of correlation between the barrier height and radius is usual (see, e.g. tables III and IV in [25]); it appears because the closer the nuclei come to each other the larger is the Coulomb repulsion which dominates in the barrier height. The M3Y DF zero-range barrier heights and radii (filled circles in figure 4 a, b) are closer to the Migdal's ones because of obvious reasons. However it was shown [47] that the zero-range M3Y forces do not reproduce the saturation properties of nuclear matter whereas the density dependent finite range M3Y forces, which result in significantly lower barriers, do so. These are the density dependent finite range M3Y forces we use in [24, 41, 44] for successful description of the above-barrier fusion excitation functions. This observation makes questionable the applicability of the Migdal DF potential for describing the nucleus-nucleus collision, and in particular the fusion process.

The barrier curvature

$$\Omega_B = \sqrt{\frac{1}{m_R}\left|\frac{d^2U}{dR^2}\right|_B} \qquad (22)$$

is another quantity that defines the fusion cross sections, e.g. in the single-barrier penetration model (BPM). Here $m_R = m_n A_P A_T/(A_P + A_T)$. The values of $\hbar\Omega_{B0}$ (for zero angular momentum) are shown in table 3 whereas the corresponding fractional differences are presented in figure 4 c). From this figure one can conclude that for the lighter colliding systems the Migdal-barriers are sharper than the M3Y-barriers whereas for the heavier systems the situation turns to the opposite.

It is interesting to compare our results concerning the system $^{16}$O+$^{208}$Pb with those obtained microscopically within the TDHF approach in Ref. [18]. In figure 11 of that work the heights of fusion barriers are shown calculated both self-consistently (accounting for the time evolution of densities) $V_B^{DD}$ and within the frozen density approximation $V_B^{FD}$. Our barrier height calculated with the M3Y density dependent forces $U_{B0\ M3Y} = 75.6$ MeV is in very good agreement with $V_B^{FD} = 76.0$ MeV. The barrier height obtained with Migdal forces $U_{B0\ MIG} = 78.4$ MeV is by approximately 2% higher than both $V_B^{FD}$ and $V_B^{DD}$. The latter varies from 75.0 up to 76.5 MeV in figure 11 of Ref. [18] in the collision energy interval considered in our work ($E_{c.m.} = 85.0 \div 109.5$ MeV). Thus, for this reaction our semi-microscopic M3Y calculations are in surprisingly good agreement with the very computer time consuming self-consistent TDHF calculations in contradistinction to the Migdal DF calculations.

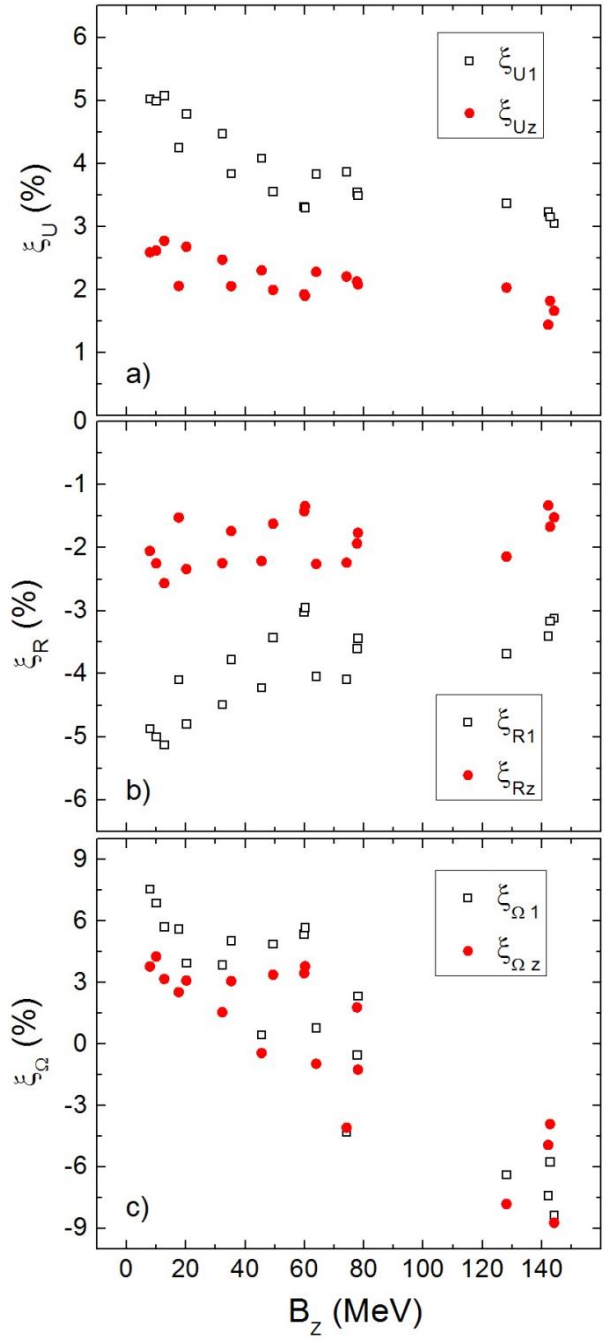

**Figure 4.** (Color online) Fractional differences for the barrier height (a), radius (b), and stiffness (frequency) (c) versus $B_Z$ (see Eqs. (19-21)). The characteristics of the Migdal DF barrier are compared with those of the M3Y DF with the zero-range exchange forces (subscript 'z') and with the density dependent ($k_{DD} = 1$) exchange forces (subscript '1').

**Table 3.** Parameters of the potentials for the reactions under consideration: the approximate Coulomb barrier height $B_Z$ (see Eq. (18)); the sum of projectile and target radii ($R_{PT0} = r_0(A_P^{1/3} + A_T^{1/3}), r_0 = 1.2$ fm); the Coulomb barrier radius $R_{B0}$, height $U_{B0}$, and curvature $\hbar\Omega_{B0}$ calculated at zero angular momentum for the M3Y zero-range (M3Yz), for the M3Y density dependent finite-range (M3Y1), and for the Migdal (MIG) interaction; the reference to the experimental fusion (capture) cross sections used in figure 5.

| Reaction | $B_Z$ (MeV) | $R_{PT0}$ (fm) | M3Yz ($k_{DD} = -1$) | | | M3Y1 ($k_{DD} = 1$) | | | MIG | | | Expt. data Refs. |
|---|---|---|---|---|---|---|---|---|---|---|---|---|
| | | | $R_{B0}$ (fm) | $U_{B0}$ (MeV) | $\hbar\Omega_{B0}$ (MeV) | $R_{B0}$ (fm) | $U_{B0}$ (MeV) | $\hbar\Omega_{B0}$ (MeV) | $R_{B0}$ (fm) | $U_{B0}$ (MeV) | $\hbar\Omega_{B0}$ (MeV) | |
| $^{12}$C+$^{12}$C | 7.86 | 5.49 | 8.05 | 5.95 | 2.80 | 8.18 | 5.86 | 2.69 | 7.80 | 6.17 | 2.91 | - |
| $^{12}$C+$^{16}$O | 9.98 | 5.77 | 8.27 | 7.74 | 2.92 | 8.40 | 7.62 | 2.84 | 8.00 | 8.02 | 3.05 | - |
| $^{12}$C+$^{32}$S | 17.57 | 6.56 | 8.78 | 14.61 | 3.47 | 8.90 | 14.42 | 3.36 | 8.55 | 15.06 | 3.56 | - |
| $^{12}$C+$^{92}$Zr | 35.27 | 8.16 | 10.07 | 32.14 | 4.42 | 10.17 | 31.81 | 4.33 | 9.80 | 33.08 | 4.56 | [4] |
| $^{12}$C+$^{144}$Sm | 49.40 | 9.04 | 10.77 | 46.78 | 5.15 | 10.86 | 46.37 | 5.07 | 10.50 | 48.08 | 5.33 | [48] |
| $^{12}$C+$^{208}$Pb | 59.89 | 9.86 | 11.51 | 58.17 | 5.59 | 11.59 | 57.71 | 5.48 | 11.25 | 59.69 | 5.79 | [49] |
| $^{12}$C+$^{204}$Pb | 60.17 | 9.81 | 11.45 | 58.46 | 5.58 | 11.53 | 57.99 | 5.47 | 11.20 | 59.97 | 5.80 | [50] |
| $^{16}$O+$^{16}$O | 12.70 | 6.05 | 10.06 | 12.70 | 3.05 | 8.61 | 9.92 | 2.97 | 8.19 | 10.45 | 3.15 | - |
| $^{16}$O+$^{28}$Si | 20.16 | 6.67 | 8.86 | 16.92 | 3.44 | 8.96 | 16.70 | 3.41 | 8.55 | 17.54 | 3.55 | - |
| $^{16}$O+$^{92}$Zr | 45.49 | 8.44 | 10.28 | 42.00 | 4.50 | 10.37 | 41.58 | 4.46 | 9.95 | 43.35 | 4.48 | [4] |
| $^{16}$O+$^{144}$Sm | 63.91 | 9.31 | 10.98 | 61.21 | 5.23 | 11.06 | 60.69 | 5.14 | 10.63 | 63.11 | 5.18 | [5] |
| $^{16}$O+$^{208}$Pb | 77.68 | 10.13 | 11.72 | 76.20 | 5.50 | 11.79 | 75.61 | 5.63 | 11.38 | 78.39 | 5.60 | [6] |
| $^{16}$O+$^{204}$Pb | 78.03 | 10.09 | 11.66 | 76.57 | 5.64 | 11.73 | 75.98 | 5.44 | 11.34 | 78.73 | 5.57 | [51] |
| $^{28}$Si+$^{28}$Si | 32.27 | 7.29 | 9.21 | 28.50 | 3.82 | 9.31. | 28.18. | 3.73 | 8.91 | 29.50 | 3.88 | - |
| $^{28}$Si+$^{92}$Zr | 74.16 | 9.07 | 10.63 | 71.15 | 4.84 | 10.70 | 70.54 | 4.85 | 10.28 | 73.38 | 4.65 | [4] |
| $^{28}$Si+$^{208}$Pb | 128.1 | 10.75 | 12.06 | 129.58 | 6.07 | 12.11 | 128.74 | 5.99 | 11.68 | 133.23 | 5.63 | [52] |
| $^{36}$S+ $^{208}$Pb | 142.19 | 10.92 | 12.40 | 144.21 | 5.53 | 12.24 | 145.53 | 6.08 | 12.05 | 148.06 | 5.27 | [53] |
| $^{36}$S+ $^{204}$Pb | 142.78 | 11.07 | 12.34 | 144.86 | 5.58 | 12.46 | 143.28 | 5.66 | 12.01 | 148.63 | 5.37 | [54] |
| $^{32}$S+$^{208}$Pb | 144.18 | 11.03 | 12.18 | 146.48 | 6.10 | 12.39 | 143.94 | 5.68 | 11.87 | 150.11 | 5.61 | [52] |

## 4. Comparison with the data

*4.1 Calculation of fusion cross sections*

It was proved that accounting for coupling to the collective modes in the target and projectile nuclei is vital for describing the modern precision experimental data on the heavy-ion capture cross sections at the near- and below-barrier energies [2, 55]. Yet at the well-above-barrier energies these couplings became unimportant, and the data could be analyzed within the framework of the BPM [3] or of the trajectory fluctuation-dissipation model, TMSF, [41, 44].

In both cases the capture (fusion) cross sections $\sigma_{th}$ is calculated according to the standard formula:

$$\sigma_{th} = \frac{\pi\hbar^2}{2m_R E_{c.m.}} \sum_{L=0}^{\infty} (2L+1) T_L. \tag{23}$$

In the parabolic approximations within the BPM approach the transmission coefficients $T_L$ are evaluated using the well-known formula:

$$T_L = \{1 + \exp[2\pi (U_{BL} - E_{c.m.})/(\hbar\Omega_{BL})]\}^{-1}. \quad (24)$$

Within the TMSF approach the calculation of $T_L$ is more involved. First, the dynamical evolution of the collective momentum $p$ and its conjugate coordinate $R$ corresponding to the radial motion is modeled using the stochastic equations with the Gaussian noise and instant friction (see details in Refs. [33, 41, 44]):

$$\frac{dp}{dt} = -\frac{dU}{dR} + \frac{\hbar^2 L^2}{m_R R^3} - K_R \frac{p}{m_R}\left(\frac{dU_n}{dR}\right)^2 + b\sqrt{2\theta K_R \left(\frac{dU_n}{dR}\right)^2} ; \quad (25)$$

$$\frac{dq}{dt} = \frac{p}{m_R} \quad (26)$$

Here the surface friction expressions [11, 12] for the dissipation and fluctuation forces are used. $K_R$ denotes the dissipation strength coefficient, $\theta$ stands for the temperature which defines the amplitude of fluctuations according to the Einstein relation.

In the TMSF, typically $20 \cdot (2L+1)$ trajectories were simulated for every partial wave until the number of captured trajectories for the particular $L$ becomes zero (see details in Refs. [34, 41, 44]). The transmission coefficient is defined as the ratio of the captured trajectories number to the full number of trajectories for particular $L$-value. The capture conditions were as described in Sec. II F of Ref. [24].

*4.2 Results*

In figure 5, three sorts of calculations are compared with the experimental above-barrier fusion cross sections, $\sigma_{exp}$. Here the ratio $r_\sigma = \sigma_{th} / \sigma_{exp}$ is shown versus the ratio $U_{B0}/E_{c.m.}$. The triangles correspond to the BPM calculations (Eqs. (23), (24)) which are performed with the Migdal (triangles down) and M3Y (triangles up) DF potentials. The circles represent the dynamical TMSF calculations (Eqs. (23), (25), (26)) with the M3Y DF potential. The comparison is performed for 12 reactions induced by $^{12}$C, $^{16}$O, $^{28}$Si, $^{32}$S, and $^{36}$S. Typical experimental errors of the data used for this figure vary from 0.5% (this is within the symbol size) up to 5% which is half width of the stripe around unity indicated in the figure. More detailed presentation of the errors can be found in figure 12 of Ref. [44]. Some reactions, for which the barrier characteristics are listed in table 3, are not included into this analysis because the data are uncertain (see [56]), not accurate enough or absent.

We see in figure 5 that the M3Y BPM points (triangles up) are always higher than unity; in many cases they are significantly higher. This means that the corresponding theoretical cross sections are higher (or much higher) than the experimental ones. No definite tendency with the energy can be seen in these points. The MIG BPM points (triangles down) behave in absolutely different way: they mostly lay significantly below unity approaching it sometimes at higher energies.

For the given nucleus-nucleus potential, the TMSF always results in the cross sections which are lower than the BPM ones due to accounting for the energy dissipation (see e.g. figure 9b of Ref. [24]). The BPM approach provides the upper limit which can be reached within TMSF when the friction coefficient is close to zero. Varying the dissipation strength coefficient $K_R$ in Eq. (25) in order to find the best agreement with the data, we had obtained in Ref. [41] the results presented in figure 5 as M3Y TMSF (circles). Indeed the circles mostly lay within the 5%-stripe.

It is interesting to note that in Ref. [56], for the reaction $^{16}$O+$^{16}$O, our cross sections calculated with M3Y DF potential appeared in good agreement with the time dependent Hartree-Fock calculation of Ref. [58]. For this reaction the results obtained within BPM and TMSF are close to each other since the reaction is light and therefore the capture happens at large center-to-center distance for which the friction form-factor in Eq. (25) is small.

Turning back to figure 5, we have to stress that the Migdal DF potential leaves no room for friction because the upper limit of the possible dynamical calculations (BPM) is already significantly lower than the experimental data.

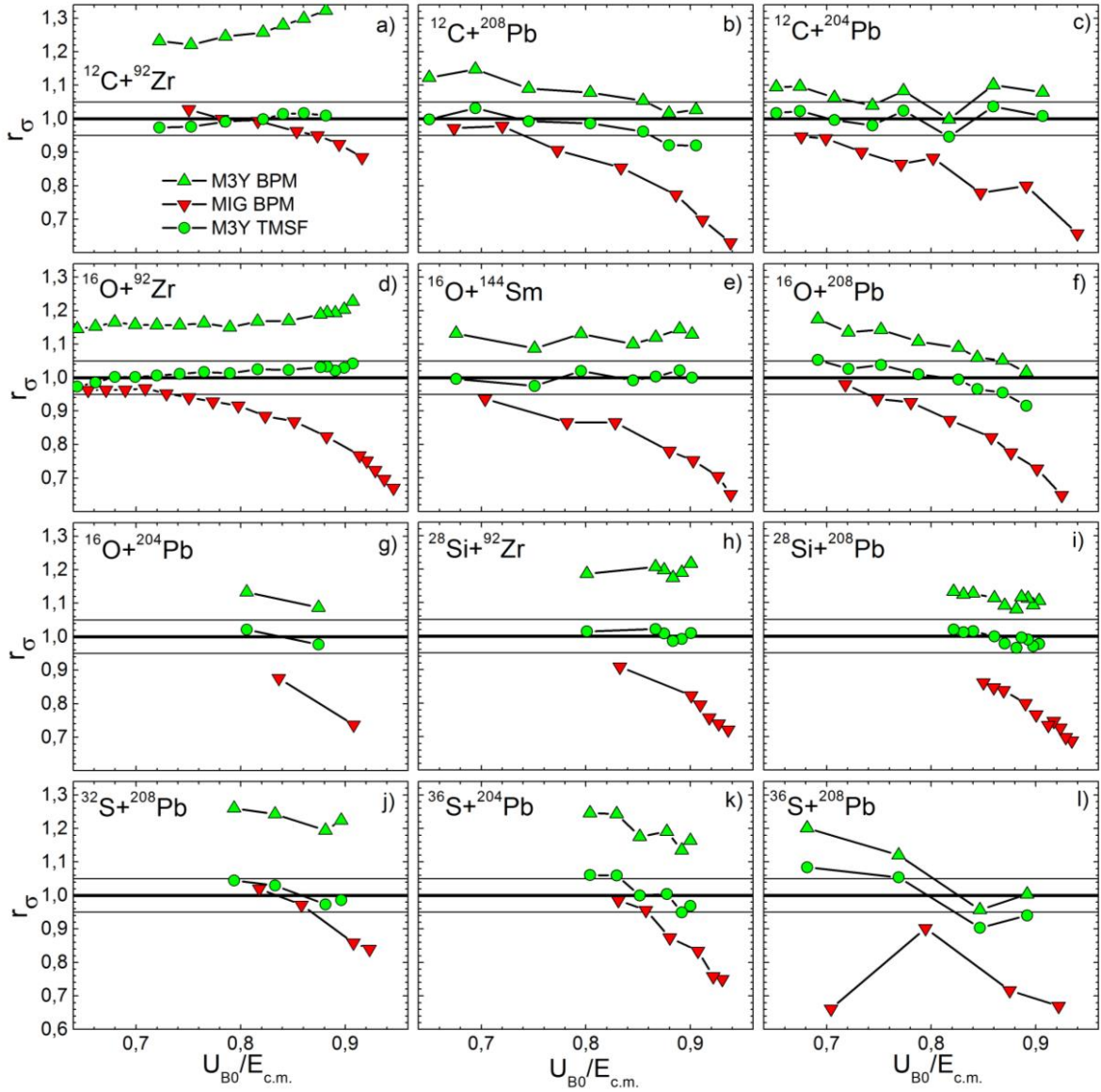

**Figure 5.** (Color online) The ratio $r_\sigma = \sigma_{th}/\sigma_{exp}$ as the function of $U_{B0}/E_{c.m.}$ for 12 reactions listed in table 3. Three versions of calculations are presented: the BPM calculations with Migdal (MIG BPM) and M3Y (M3Y BPM) DF potentials as well as the dynamical calculations [41] with the M3Y DF potential (M3Y TMSF). The references to the experimental data are given in table 3, in many cases the data are taken from [57].

## 5. Conclusions

In Refs. [41, 44] we calculated the heavy ion fusion (capture) excitation functions for reactions involving spherical nuclei and obtained good agreement with the data for the above barrier region. Those calculations were performed within the framework of the trajectory fluctuation-dissipation model using the double-folding nucleus-nucleus potential with M3Y $NN$ forces. Another option of the double folding model, namely the one with the Migdal $NN$ forces, often was used in the literature. The attractive feature of the Migdal forces is their zero-range that makes the double folding calculations easy.

In the present work we systematically compared the Coulomb barriers obtained both with the M3Y and Migdal forces. The input of these calculations was otherwise the same. In the calculations we used the nuclear densities resulted from the Hartree-Fock approach with the SKX-Skyrme forces accounting for the tensor part. The charge density distributions obtained in those calculations were in good agreement with the experimental data. Thus one could hope that the proton and neutron density distributions obtained in the same calculations represented the real situation as well.

The comparison made for 19 reactions with spherical nuclei revealed that the Migdal barriers are always higher by several percent than the M3Y barriers, even when the zero-range exchange forces were used in the M3Y calculations. These relatively high Migdal barriers result in rather low fusion cross sections calculated within the single-barrier penetration model: most of the calculated cross sections significantly (up to 40%) underestimates the experimental values.

An attempt to apply the Migdal double-folding potential in the trajectory fluctuation-dissipation model inevitably results in even worse agreement with the experiment.

**Acknowledgment**

M.V.C. is grateful to the Dmitry Zimin "Dynasty" Foundation (Russia) for the financial support within the framework of individual scholarship.